\documentclass[iop,onecolumn]{emulateapj}
\usepackage{graphicx}
\usepackage{multirow}
\usepackage{textcomp}
\usepackage{epsfig}
\usepackage{amsmath}
\usepackage{amssymb}
\usepackage{amsthm}
\usepackage{xy}

\shorttitle{Pulse profiles from neutron stars}
\shortauthors{Turolla \& Nobili}

\begin{document}

\title{Pulse profiles from thermally emitting neutron stars}

\author{R. Turolla\altaffilmark{1} and L. Nobili\altaffilmark{*}}
\affil{Department of Physics and Astronomy, University
of Padova, Via Marzolo 8, I-35131 Padova, Italy}
\altaffiltext{1}{Also at Mullard Space Science Laboratory, University
College London, Holmbury St. Mary, Dorking, Surrey, RH5 6NT, UK}
\altaffiltext{*}{Retired}

\begin{abstract}

The problem of computing the pulse profiles from thermally emitting spots on the
surface of a neutron star in general relativity is reconsidered. We show that it is possible to extend
\cite{belob02} approach to include (multiple) spots of finite size in different positions on the star surface.
Results for the pulse profiles are expressed by comparatively simple analytical formulas which involve only
elementary functions.

\end{abstract}


\section{Introduction}\label{intro}

X-ray emission from isolated neutron stars (NSs), first detected
in radio pulsars (PSRs), is now increasingly observed in other
classes of sources, most of which are radio-silent or have radio
properties much at variance with those of PSRs. They include the
thermally emitting NSs \cite[XDINSs; e.g.][]{turolla09}, the
central compact objects in supernova remnants \cite[CCOs;
e.g][]{deluca08}, the magnetar candidates \cite[SGRs and AXPs;
e.g.][]{mereghetti08,reaesp11} and the rotating radio transients
\cite[RRaTs; e.g.][]{buspo12}.

With the exception of some PSRs, the X-ray emission of which is
dominated by a non-thermal component of magnetospheric origin, the
spectra of all other X-ray emitting, isolated NSs exhibit one (or more)
thermal component which, most probably, originates at the star
surface. Since pulsations are observed, thermal X-ray photons come
either from a localized, heated region, like in SGRs/AXPs and
PSRs, or from the entire cooling surface with an inhomogeneous
temperature distribution, like in XDINSs. In this respect the
analysis of the observed pulse profiles in different energy
bands is bound to reveal much on the surface thermal map of the
NS, on the physical size and position of the emitting regions(s)
in particular \cite[e.g.][]{zt06,alba10}.

The problem of modelling the pulse profiles of a rotating, thermally emitting
NS, including the effects of gravitational ray bending, is an old
one and has been thoroughly addressed in the literature
\cite[e.g.][]{pfc83,page95,pagesar96,pod00,belob02,zt06}. In
particular, in their classic paper \cite{pfc83} analyzed the
emission from two antipodal, uniform, circular caps. Although
their approach contains no inherent complexity, the treatment of
photon propagation in a Schwarzschild spacetime leads to elliptic
integrals and requires numerical evaluation. In general, resorting
to a numerical approach is unavoidable every time a continuous
surface temperature distribution, anisotropic emission and/or an
arbitrary shape of the emission regions have to be accounted for.
However, \cite{belob02}, by means of a clever approximation, has
shown that simple, analytical expressions can be derived for the
pulse profiles in full general relativity (Schwarzschild spacetime) for
point-like spots.

In this paper we make use of \cite{belob02} approximate treatment
to extend his analysis to the case of finite, uniform, circular
spots. Our results are valid for an arbitrary number of spots,
regardless of their size, temperature and mutual position on the
star surface (e.g. two different, non-antipodal caps). Some more
complex emission geometries (like a cap surrounded by a corona)
can also be easily accommodated. The expression for the total
observed flux is analytical and this makes our approach both
simple and fast for the evaluation and comparison of pulse profiles with
observations.

\section{Observed flux}\label{obsflux}

Let us consider a surface element $dS$ on a neutron star of radius
$R$ and mass $M$ and let us assume that the Schwarzschild solution
correctly describes the spacetime outside the star (in the
following $R_S=2GM/c^2$ is
the Schwarzschild radius). Let us further introduce a
spherical coordinate system, $(r,\, \theta,\, \phi)$, centered on
the star in such a way that the polar axis coincides with the
line-of-sight (LOS; unit vector $\mathbf{\hat l}$). The distance
to the observer is $D \gg R$.

\begin{figure}
\epsscale{0.35}
\plotone{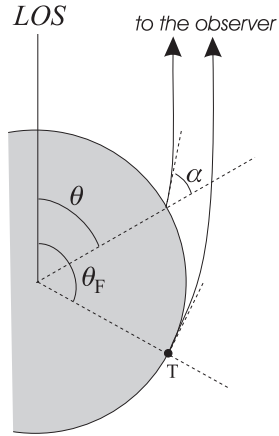} \caption{
A schematic view of ray propagation. The angles $\theta$, $\alpha$ and $\theta_F$ are also shown.
\label{rays}}
\end{figure}

Because photon trajectories are not straight lines, the ray from
$dS=R^2\sin\theta d\theta d\phi$ which reaches the observer leaves the surface, with respect to
the local normal, at an angle $\alpha\neq\theta$ (see Figure \ref{rays}). The relation between
$\alpha$ and $\theta$ is given, implicitly, by the two equations

\begin{equation}\label{theta}
\theta=\int_R^\infty\frac{dr}{r^2}\left[\frac{1}{b^2}-\frac{1}{r^2}\left(
1-\frac{R_S}{r}\right)\right]^{-1/2}
\end{equation}

\begin{equation}\label{alpha}
\sin\alpha=\frac{b}{R}\left(1-\frac{R_S}{R}\right)^{1/2}\,,
\end{equation}
where $b$ is the ray impact parameter \cite[][]{belob02}.

The (monochromatic) flux $dF_\nu$ from $dS$ detected by the observer is then

\begin{equation}
dF_\nu=\left(1-\frac{R_S}{R}\right)I_\nu\cos\alpha
\left(\frac{d\cos\alpha}{d\cos\theta}\right)\frac{dS}{D^2}\,.
\end{equation}
where $\nu$ is the photon frequency and $I_\nu$ the specific
intensity, both measured by the static observer at $r=R$. The total flux
is obtained by
integrating the previous expression over the visible part of the emitting
region, $S_V$. If the emission is Planckian at the local (uniform) temperature $T$, $I_\nu=B_\nu(T)$
and this results in

\begin{equation}\label{flux}
F_\nu=\left(1-\frac{R_S}{R}\right)B_\nu(T)\int_{S_V}\cos\alpha
\left(\frac{d\cos\alpha}{d\cos\theta}\right)\frac{dS}{D^2}\,.
\end{equation}
In Newtonian gravity it is $\alpha=\theta$ and the flux is simply
proportional to the area of the visible emitting region projected in the
plane of the sky.

\cite{belob02} found that a simple, approximate expression can be used to
link $\alpha$ and $\theta$, without the need to solve (numerically) eqs.
(\ref{theta}) and (\ref{alpha}),

\begin{equation}\label{cosalp}
1-\cos\alpha=(1-\cos\theta)\left(1-\frac{R_S}{R}\right)\,.
\end{equation}
Eq. (\ref{cosalp}) is remarkably accurate and produces a
fractional error $\lesssim 3\%$ for $R\gtrsim 3R_S$. Substituting
$\cos\alpha$ and $d\cos\alpha/d\cos\theta$ into eq. ({\ref{flux}),
one obtains

\begin{equation}\label{flux1}
F_\nu=\left(1-\frac{R_S}{R}\right)^2B_\nu(T)\int_{S_V}
\left[\frac{R_S}{R}+\left(1-\frac{R_S}{R}\right)\cos\theta\right]\frac{dS}{D^2}\,.
\end{equation}
The flux is, then, expressed by the sum of two contributions, the
first proportional to the surface area and the second to the
projected area of the visible part of the emitting region. The latter, apart for the
factor $(1-R/R_S)$, is the analogue of the Newtonian expression, while the former
is a purely relativistic correction.
The problem of
computing the flux, once the geometry is fixed, is therefore
reduced to that of determining $S_V$ and evaluating the two
integrals
\begin{equation}\label{integs}
I_p=\int_{S_V}\cos\theta \sin\theta\,d\theta d\phi\,,\quad
I_s=\int_{S_V}\sin\theta\,d\theta d\phi\,.
\end{equation}

\subsection{Single circular spot}\label{single}

In order to proceed further, we consider first the simplest case,
in which the emitting region is a circular cap of semi-aperture
$\theta_c$ with its center at $(R,\,\theta_o,\, 0)$. For the sake
of simplicity, and also because this is the most common
occurrence, we consider only the case \footnote{Despite this
limitation many configurations of interest can be nevertheless
treated (see \S\ref{multi}).} $\theta_c\leq\pi/2$. Moreover, we
restrict to $0\leq\theta_o\leq\pi$, since the case
$\pi\leq\theta_o\leq 2\pi$ is reduced to the previous one upon the
substitution $\theta_o\to 2\pi-\theta_o$, given the axial symmetry
around the LOS.

The $\phi$-integral in both $I_p$ and $I_s$ is immediate. By
denoting with $\phi_b(\theta)$ the cap boundary
($0\leq\phi_b\leq\pi$), it is

\begin{equation}\label{integs1}
I_p=2\int_{\theta_{\mathrm{min}}}^{\theta_{\mathrm{max}}}\cos\theta
\sin\theta\phi_b(\theta)\,d\theta\,,
\quad I_s=2\int_{\theta_{\mathrm{min}}}^{\theta_{\mathrm{max}}}\sin\theta\phi_b(\theta)\,d\theta\,,
\end{equation}
where $\theta_{\mathrm{min}}$, $\theta_{\mathrm{max}}$ are the limiting values of
the co-latitude, which are discussed below.

\begin{figure}
\epsscale{0.75}
\plottwo{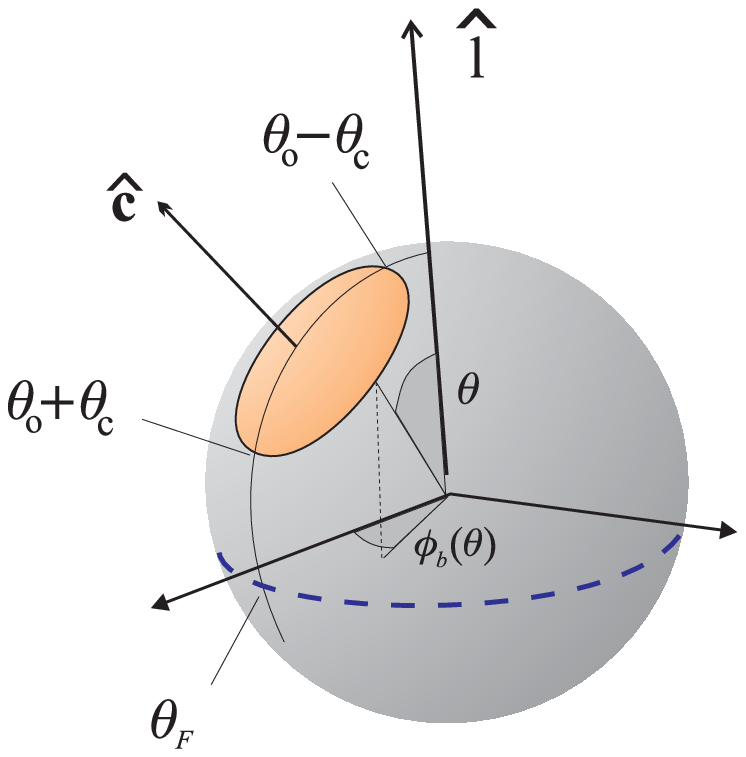}{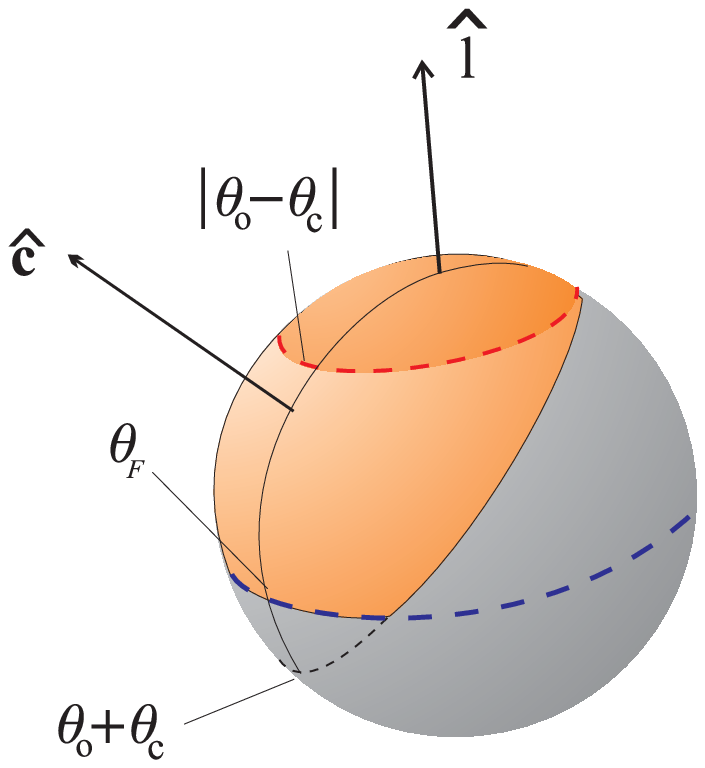}
\caption{ Geometry
for a single spot; the visible part is in orange and the
terminator is marked by the blue, dashed line. (\emph{Left}) The
spot is fully into view and $\theta_o-\theta_c>0$. (\emph{Right})
The spot is partially into view and $\theta_o-\theta_c<0$; the
darker part bounded by the dashed orange line is the region
$0\leq\theta\leq\vert\theta_o-\theta_c\vert$ (see text).
\label{caps}}
\end{figure}

The function $\phi_b$ can be readily found noticing that a generic
point on the cap boundary has coordinates $(R\,,\theta,\, \phi_b)$.
In  a spherical coordinate system with the polar axis coincident
with the cap axis (unit vector $\mathbf{\hat c}$), its coordinates
are $(R\,, \Theta\equiv\theta_c,\, \Phi)$; the latter system is rotated by an angle $\theta_o$
with respect to the former around an axis perpendicular to the $\mathbf{\hat
l}$--$\mathbf{\hat c}$ plane. By exploiting the transformation between the (cartesian) coordinates in the
two systems, one gets

\begin{equation}\label{transf1}
\cos\theta=\cos\theta_o\cos\theta_c-\sin\theta_o\sin\theta_c\cos\Phi\,,
\end{equation}

\begin{equation}\label{transf2}
\sin\theta\cos\phi_b=\cos\theta_o\sin\theta_c\cos\Phi+\sin\theta_o\cos\theta_c\,.
\end{equation}
Solving the second for $\cos\Phi$ and substituting into the
first one, one finally obtains

\begin{equation}\label{phic}
\cos\phi_b=\frac{\cos\theta_c-\cos\theta_o\cos\theta}{\sin\theta_o\sin\theta}\,;
\end{equation}
$\cos\phi_b$ is actually related to the function $h$ introduced by
\cite{pfc83}. It is immediate to verify that it is $-1\leq
\cos\phi_b\leq 1$, and hence $\phi_b$ is defined, in the
range
$\vert\theta_o-\theta_c\vert<\theta<\min{[\theta_o+\theta_c,
2\pi-(\theta_o+\theta_c)]}$, since it must be, by definition,
$0\leq\theta\leq \pi$. It is important to notice that the
$\theta$-range in which it is possible to define $\cos\phi_b$ is
not sufficient to cover the entire cap when the LOS intersects the
cap itself: this occurs either for $\theta_o-\theta_c<0$ or
$\theta_o+\theta_c>\pi$. Since it is $0<\theta_c\leq\pi/2$ by
assumption and disregarding visibility, there is just one
intersection, at $\theta=0$ or $\pi$. In these cases the cap is
fully covered only by adding the range
$0\leq\theta\leq\vert\theta_o-\theta_c\vert$ or
$2\pi-(\theta_o+\theta_c)\leq\theta\leq\pi$, respectively. This
accounts for the missing surface which is itself a circular cap,
perpendicular to the LOS, where $\phi$ spans the entire range
$[0\,,2\pi]$ (see fig. \ref{caps}). Accordingly, the definition of
$\phi_b(\theta)$ can be continuously extended as

\begin{equation}\label{phicext}
\phi_b(\theta)=
\begin{cases}
\arccos{\displaystyle\left[\frac{\cos\theta_c-\cos\theta_o\cos\theta}{\sin\theta_o\sin\theta}\right]} &
\vert\theta_o-\theta_c\vert<\theta<\min{[\theta_o+\theta_c, 2\pi-(\theta_o+\theta_c)]}  \cr
\cr
\pi & 0\leq\theta\leq\vert\theta_o-\theta_c\vert \quad \mathrm{if}\ \theta_o-\theta_c\leq0\cr
& 2\pi-(\theta_o+\theta_c)\leq\theta\leq\pi \quad \mathrm{if}\ \theta_o+\theta_c\geq\pi \cr
\end{cases}
\end{equation}
to include all cases.

At each visible point of the star surface it has to be $\cos\alpha\geq 0$, and the
terminator lies precisely at $\alpha=\pi/2$. From eq. (\ref{cosalp})
it follows that the terminator co-latitude is given by

\begin{equation}\label{termin}
\cos\theta_F=\left(1-\frac{R}{R_S}\right)^{-1};
\end{equation}
it is $\cos\theta_F\leq 0$ and $\theta_F\geq\pi/2$, as expected,
since relativistic effects bring more than half the sphere into
view.

In the case the cap is entirely visible, i.e. it does not
intersect the terminator, its co-latitude is in the
range\footnote{Note, however, that it cannot be
$\theta_{\mathrm{min}}=0$ and
$\theta_{\mathrm{max}}=2\pi-(\theta_o+\theta_c)$ because this
would imply $\theta_c>\pi/2$.}
$\theta_{\mathrm{min}}=\max{(0,\theta_o-\theta_c)}\leq\theta\leq
\theta_{\mathrm{max}}=\min{[2\pi-(\theta_o+\theta_c),\theta_o+\theta_c]}$.
The presence of the terminator (at $\pi/2\leq\theta_F<\pi$) is
easily accounted for by replacing $\theta_{\mathrm{min}}$
($\theta_{\mathrm{max}}$), as defined above, with $\theta_F$ every
time it is $\theta_F<\theta_{\mathrm{min}}$
($\theta_F<\theta_{\mathrm{max}}$). Summarizing, it is

\begin{equation}\label{thmin}
\theta_{\mathrm{min}}=\min{[\max{(0,\theta_o-\theta_c)},\theta_F]}
\end{equation}
and

\begin{equation}\label{thmax}
\theta_{\mathrm{max}}=\min{(\theta_0+\theta_c,\theta_F)}\,.
\end{equation}

Turning to the evaluation of $I_{p,s}$, both integrals become
trivial for $\phi_b=\pi$ and yield

\begin{equation}\label{phicpi}
I_p=2\int_{\theta_1}^{\theta_2}\pi\cos\theta\sin\theta\,d\theta=\pi(\sin^2\theta_2-\sin^2\theta_1)\,,\quad
I_s=2\int_{\theta_1}^{\theta_2}\pi\sin\theta\,d\theta=2\pi(\cos\theta_2-\cos\theta_1)
\end{equation}
for any pair of angles $\theta_1\,,\theta_2$. In the opposite
case, the two indefinite integrals

\begin{equation}\label{integpind}
I_1=2\int\cos\theta
\sin\theta\arccos\left[\frac{\cos\theta_c-\cos\theta_o\cos\theta}{\sin\theta_o\sin\theta}\right]\,d\theta\,,
\end{equation}
and

\begin{equation}\label{integsind}
I_2=2\int
\sin\theta\arccos\left[\frac{\cos\theta_c-\cos\theta_o\cos\theta}{\sin\theta_o\sin\theta}\right]\,d\theta
\end{equation}
have to be calculated. It turns out that $I_{1,2}$ can be evaluated analytically in terms of elementary
functions (see the Appendix for more details)

\begin{eqnarray}\label{integpindex}
I_1&=&\sin^2\theta\arccos\left[\frac{\cos\theta_c-\cos\theta_o\cos\theta}{\sin\theta_o\sin\theta}\right]
-\sin^2\theta_c\cos\theta_o\arcsin\left[\frac{\cos\theta-\cos\theta_o\cos\theta_c}{\sin\theta_o\sin\theta_c}\right]\cr
&&-\cos\theta_c\sqrt{-\left[\cos\theta-\cos(\theta_o+\theta_c)\right]\left[\cos\theta-\cos(\theta_o-\theta_c)\right]}
\end{eqnarray}
and

\begin{eqnarray}\label{integsindex}
I_2&=&-2\cos\theta\arccos\left[\frac{\cos\theta_c-\cos\theta_o\cos\theta}{\sin\theta_o\sin\theta}\right]
+2\cos\theta_c\arcsin\left[\frac{\cos\theta-\cos\theta_o\cos\theta_c}{\sin\theta_o\sin\theta_c}\right]\cr
&&+\mathrm{sign}\left(\theta_o+\theta_c-\pi\right)\arcsin\left[\frac{(\cos\theta_o\cos\theta_c+1)\cos\theta+
\sin^2\theta_o-\cos^2\theta_c-\cos\theta_o\cos\theta_c}{(1+\cos\theta)\vert\sin\theta_o\sin\theta_c\vert}\right]\cr
&&-\mathrm{sign}\left(\theta_o-\theta_c\right)\arcsin\left[\frac{(\cos\theta_o\cos\theta_c-1)\cos\theta+
\sin^2\theta_o-\cos^2\theta_c-\cos\theta_o\cos\theta_c}{(1-\cos\theta)\vert\sin\theta_o\sin\theta_c\vert}\right]\,,
\end{eqnarray}
where the arbitrary constant was set to zero. It is then $I_{p,s}=I_{1,2}(\theta_{max})-I_{1,2}(\theta_{min})$.
We note that
$I_{1,2}$ take a simple form for $\theta=\theta_o\pm\theta_c$. In
particular, if the cap is fully into view (see Figure \ref{caps},
left), it is $I_{p}=\pi\cos\theta_o\sin^2\theta_c$ and
$I_{s}=2\pi(1-\cos\theta_c)$, as it follows also from geometrical
considerations. The complete form of eqs. (\ref{integpindex}) and
(\ref{integsindex}) is actually required only when evaluating the
integrals at $\theta_F$.

The flux (eq. [\ref{flux1}]) is finally written as

\begin{equation}\label{fluxfin}
F_\nu=\left(1-\frac{R_S}{R}\right)^2B_\nu(T)A_{\mathrm{eff}}(\theta_c,\theta_o)\,,
\end{equation}
where we introduced the ``effective'' area
\begin{equation}\label{effarea}
A_{\mathrm{eff}}(\theta_c,\theta_o)=R^2\left[\frac{R_S}{R}I_s+
\left(1-\frac{R_S}{R}\right) I_p\right]\,.
\end{equation}

\subsection{Multiple spots and other geometries}\label{multi}

Having computed the flux seen by a distant observer for a single
circular spot, it is straightforward to generalize the result to
an arbitrary number of spots. We stress that this is possible
because, using Beloborodov's approximation, the flux is
proportional to the ``effective'' area of the cap, $A_{\mathrm
{eff}}$, introduced in the previous section. Although we impose no
restrictions on the parameters, the assumption that the spots do
not intersect is understood. For the sake of simplicity, here we
consider just two circular, uniform caps with semi-aperture
$\theta_{c,i}$, and temperature $T_i$ ($i=1,\, 2$). Let us further
assume that the spots are aligned, in the sense that their centers
lie on the same meridian (the more general case of misaligned caps
will be discussed in the next section), and let $\theta_{d,2}$ be
the relative angular displacement of second one with respect to
the first, the center of which is at $\theta_o$ (of course it is
$\theta_{d,1}=0$). The spot centers are then at
$\theta_{o,i}=\theta_o+\theta_{d,i}$ and the total flux can be
calculated by simply adding the two contributions

\begin{equation}\label{twocap}
F_\nu^{\mathrm{TOT}}=\left(1-\frac{R_S}{R}\right)^2\left[B_\nu(T_1)A_{\mathrm{eff}}(\theta_{c,1},\theta_o)+
B_\nu(T_2)A_{\mathrm{eff}}(\theta_{c,2},\theta_o+\theta_{d,2})\right]\,.
\end{equation}

The case of a two-temperature cap, i.e. a cap at $T_1$ surrounded by a circular corona at $T_2$, is
treated much in
the same way by subtracting from the larger cap the contribution of the inner spot and adding the latter at the
proper temperature

\begin{equation}\label{capcor}
F_\nu^{\mathrm{TOT}}=\left(1-\frac{R_S}{R}\right)^2\left\{B_\nu(T_2)\left[
A_{\mathrm{eff}}(\theta_{c,2},\theta_o)-A_{\mathrm{eff}}(\theta_{c,1},\theta_o)\right]+B_\nu(T_1)
A_{\mathrm{eff}}(\theta_{c,1},\theta_o)\right\}\,.
\end{equation}

With the aid of the previous expressions more configurations can
be modelled. In particular, for a NS with a thermal map made of
two (antipodal) caps at $T_1$ while the rest of the surface is at
$T_2$, one obtains the flux by using twice eq. (\ref{capcor}) with
$\theta_{c,2}=\pi/2$, the second time replacing $\theta_o$ with
$\theta_o+\pi$, and summing the two contributions. The similar
case of a single cap at $T_1$ is handled by summing the flux given
by eq. (\ref{capcor}) with $\theta_{c,2}=\pi/2$ and that of eq.
(\ref{fluxfin}), with $\theta_{c}=\pi/2$,
$\theta_o\to\theta_o+\pi$.

\subsection{Pulse profiles}

In order to compute pulse profiles we consider first the case of a single spot. Let $\mathbf{\hat r}$ be
a unit vector parallel to the rotation axis and $\Omega$ the star angular velocity, $\Omega=2\pi/P$
where $P$ is the spin period. Observed periods in X-ray emitting INSs are in the range $\approx 0.1$--10 s, so
the assumption of Schwarzschild spacetime previously introduced is fully justified. We also
introduce the angles $\chi$, $\xi$ between the LOS, the cap axis and the rotation axis, respectively, i.e.
$\cos\chi=\mathbf{\hat r}\cdot\mathbf{\hat l}$ and $\cos\xi=\mathbf{\hat r}\cdot\mathbf{\hat c}$.

Since the cap co-rotates with the star, the vector $\mathbf{\hat c}$ rotates around $\mathbf{\hat r}$, keeping
$\xi$ constant. This implies that $\theta_o$ changes in time. Introducing the rotational phase $\gamma=\Omega
t+\gamma_0$ ($\gamma_0$ is an arbitrary initial phase), from simple geometrical considerations it follows
that

\begin{equation}\label{vgeom}
\cos\theta_o=\cos\chi\cos\xi-\sin\chi\sin\xi\cos\gamma\,.
\end{equation}
Eq. (\ref{fluxfin}) then provides the phase-resolved spectrum once the previous expression is used for
$\cos\theta_o$.
The pulse profile in a given energy band is immediately obtained integrating over frequencies. Since
$\int_0^\infty B_\nu(T)\, d\nu=\sigma T^4/\pi$ ($\sigma$ is the
Stefan-Boltzmann constant), the pulse profile in the $[\nu_1,\,\nu_2]$ range is

\begin{equation}\label{ppband}
F(\nu_1,\nu_2)=\left(1-\frac{R_S}{R}\right)^2 C(\nu_1,\nu_2)\frac{\sigma T^4}{\pi}
A_{\mathrm{eff}}(\theta_c,\theta_o)\,,
\end{equation}
where $C(\nu_1,\nu_2)=(\pi/\sigma T^4)\int_{\nu_1}^{\nu_2} B_\nu(T)\, d\nu$. Similar expressions hold
for other geometries.

\begin{figure}
\plottwo{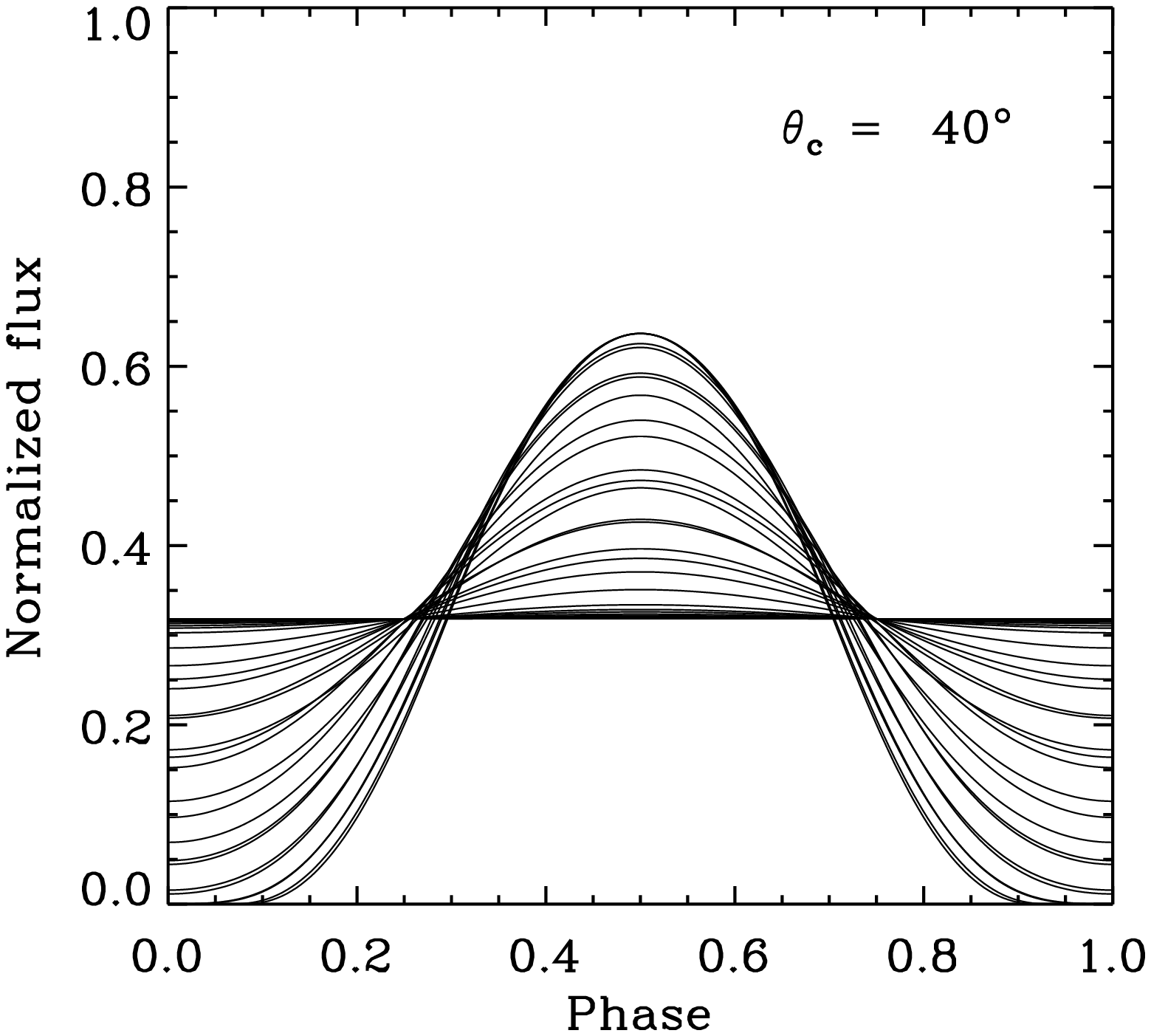}{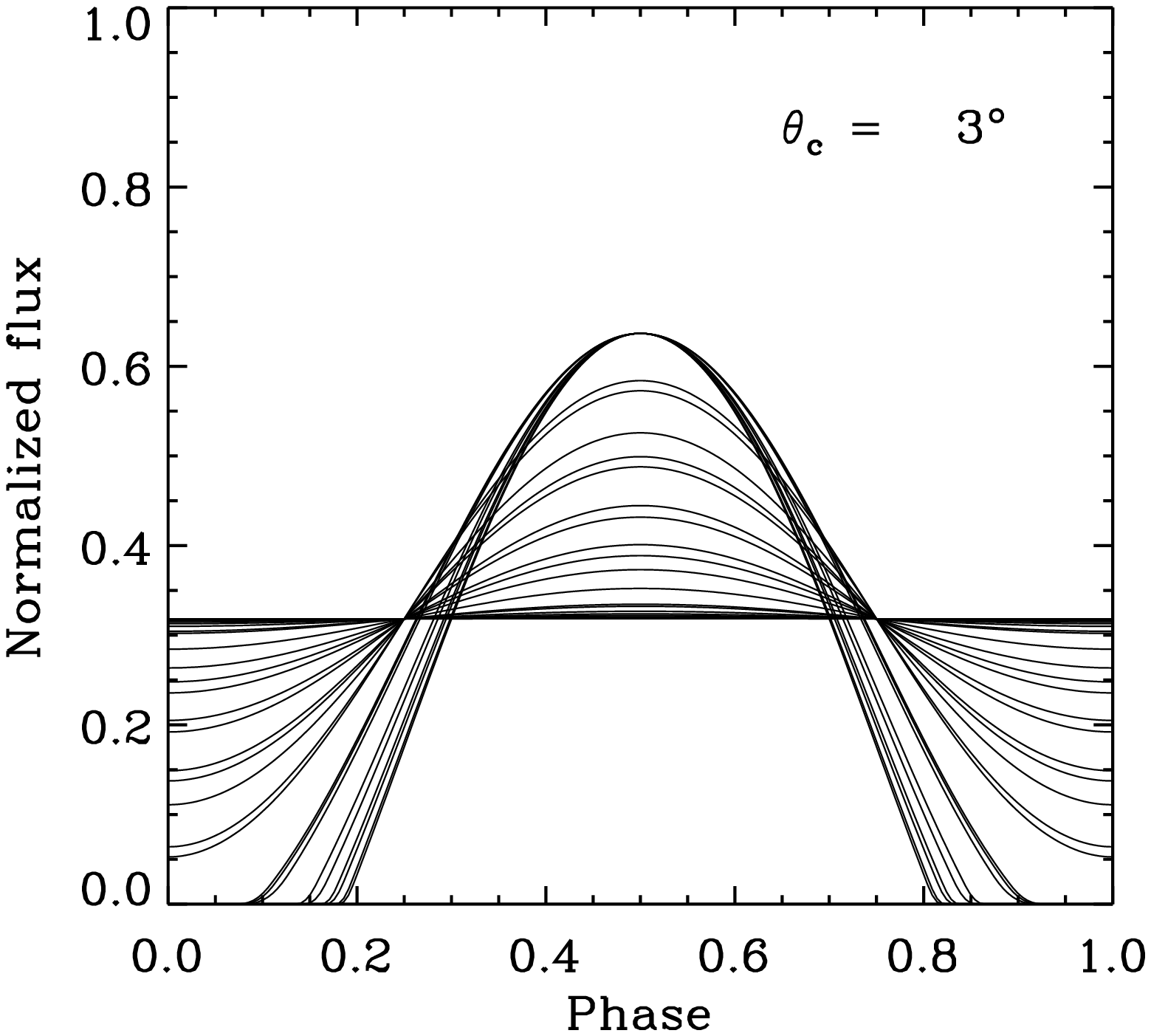} \caption{(\emph{Left})
Normalized flux as a function of phase for a single spot of
semi-aperture $\theta_c=40^\circ$ and different values of $\chi$
and $\xi$; (\emph{right}) same, but for $\theta_c=3^\circ$.
\label{pp-one}}
\end{figure}

Some examples are illustrated in Figure \ref{pp-one}, \ref{pp-two}
and \ref{pp-mis} where the pulse profiles are shown for seven
values of the angles $\chi,\,\xi$ in the range
$[0^\circ,\,90^\circ]$, step $15^\circ$. Because eq. (\ref{vgeom})
is invariant by exchanging $\chi$ and $\xi$, only the 28 pulse profiles which are actually
diverse are shown.
In all cases it is $M=1.4\, M_\odot$ and $R=15$ km ($R/R_S=3.6$),
corresponding to $\theta_F=112^\circ$, and the pulse profiles
refer to the bolometric flux (i.e. $C=1$), normalized to
$(F_{\mathrm{max}}+F_{\mathrm{min}})/2$. Figure \ref{pp-one} shows
the case of a single spot for $\theta_c=40^\circ$ (left) and
$\theta_c=3^\circ$ (right), small enough to te treated as
point-like (see \S\ref{discus}). The pulse profiles for two equal,
antipodal ($\theta_{d,2}=180^\circ$) spots are illustrated in
Figure \ref{pp-two}, again for $\theta_c=30^\circ$ (left) and
$\theta_c=3^\circ$ (right).
\begin{figure}
\plottwo{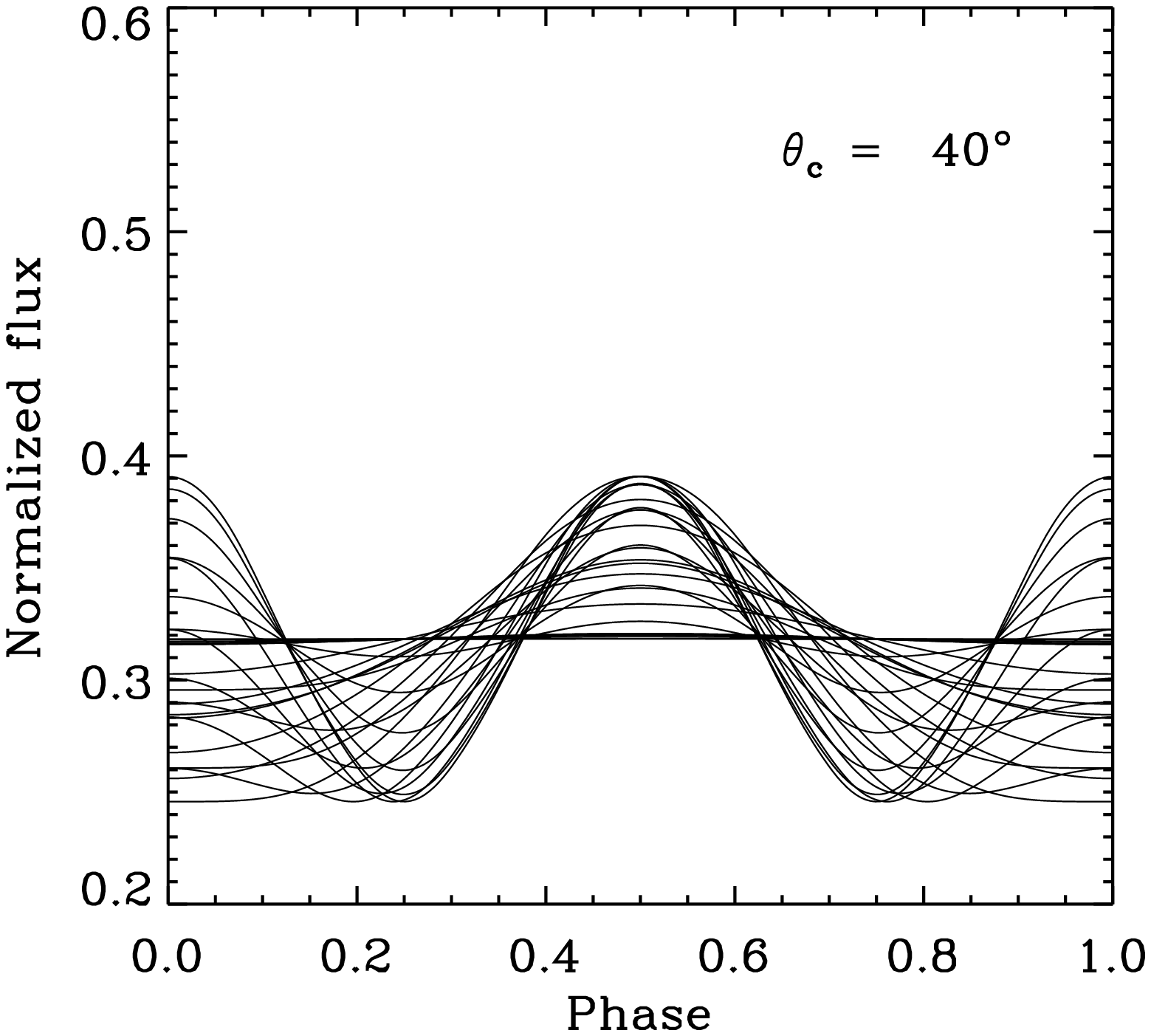}{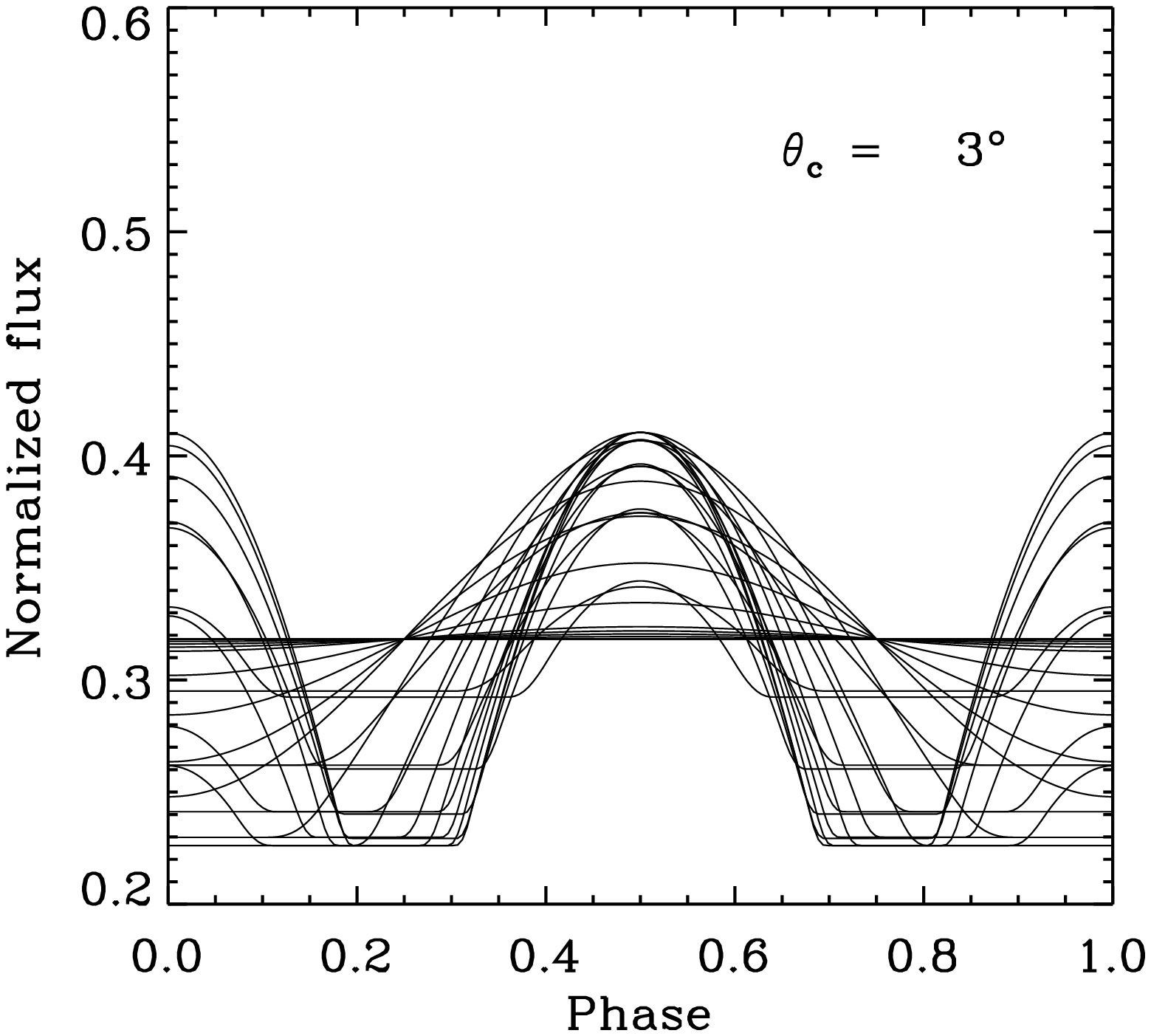}\caption{
Same as in fig. \ref{pp-one}, but for two equal, antipodal caps.
\label{pp-two}}
\end{figure}
Figure \ref{pp-mis} (left) refers
to two non-antipodal ($\theta_{d,2}=120^\circ$), different
($\theta_{c,1}=30^\circ\,,\theta_{c,2}=45^\circ$, $kT_1=0.4$ keV,
$kT_2=1$ keV) caps, while the right  panel illustrates the same case, but with the second spot
shifted in longitude by $45^\circ$. The latter is simply obtained by adding a constant phase-shift to
$\gamma$ in eq. (\ref{vgeom}) when $\theta_o$ refers to the second spot.

\begin{figure}
\plottwo{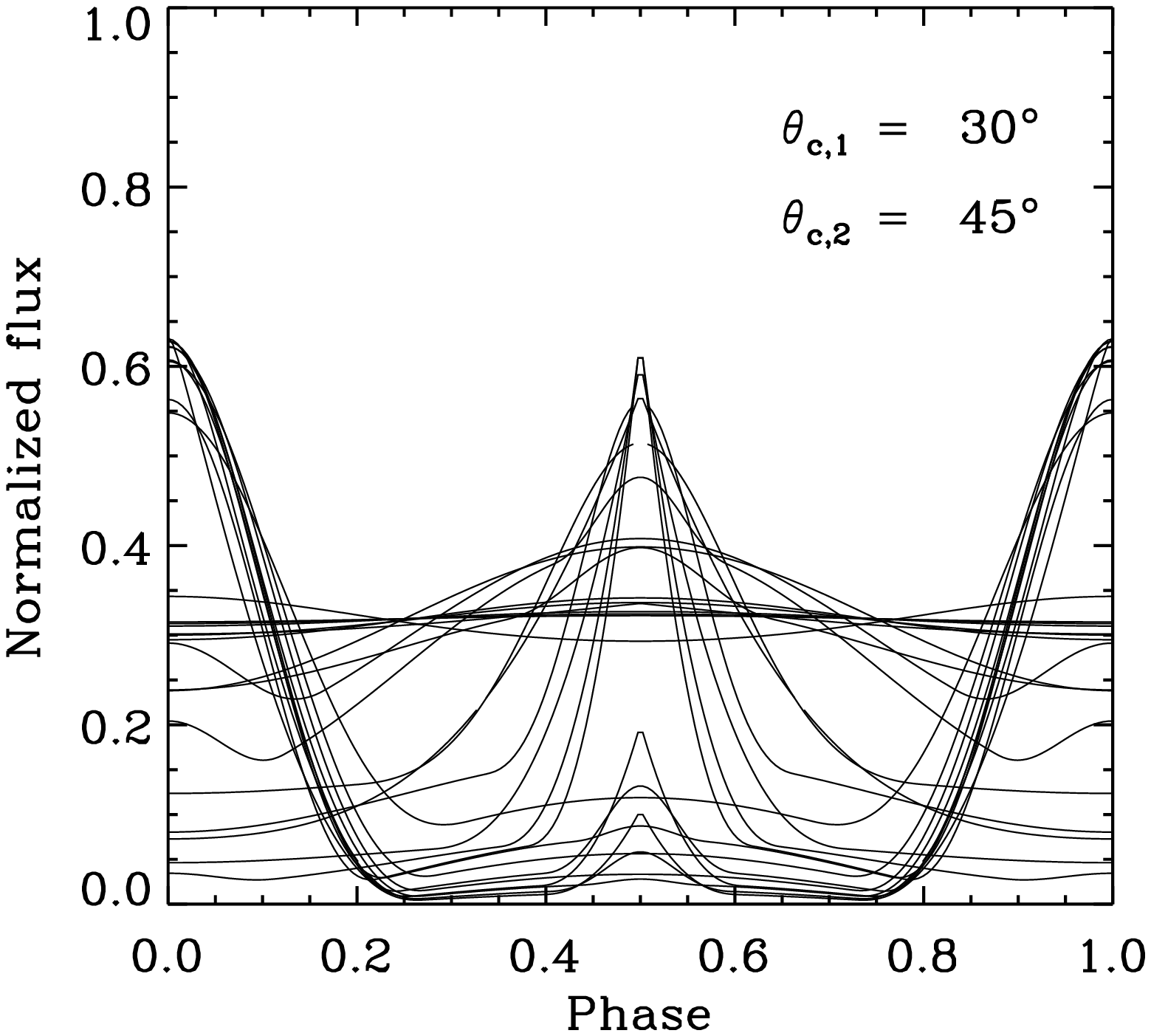}{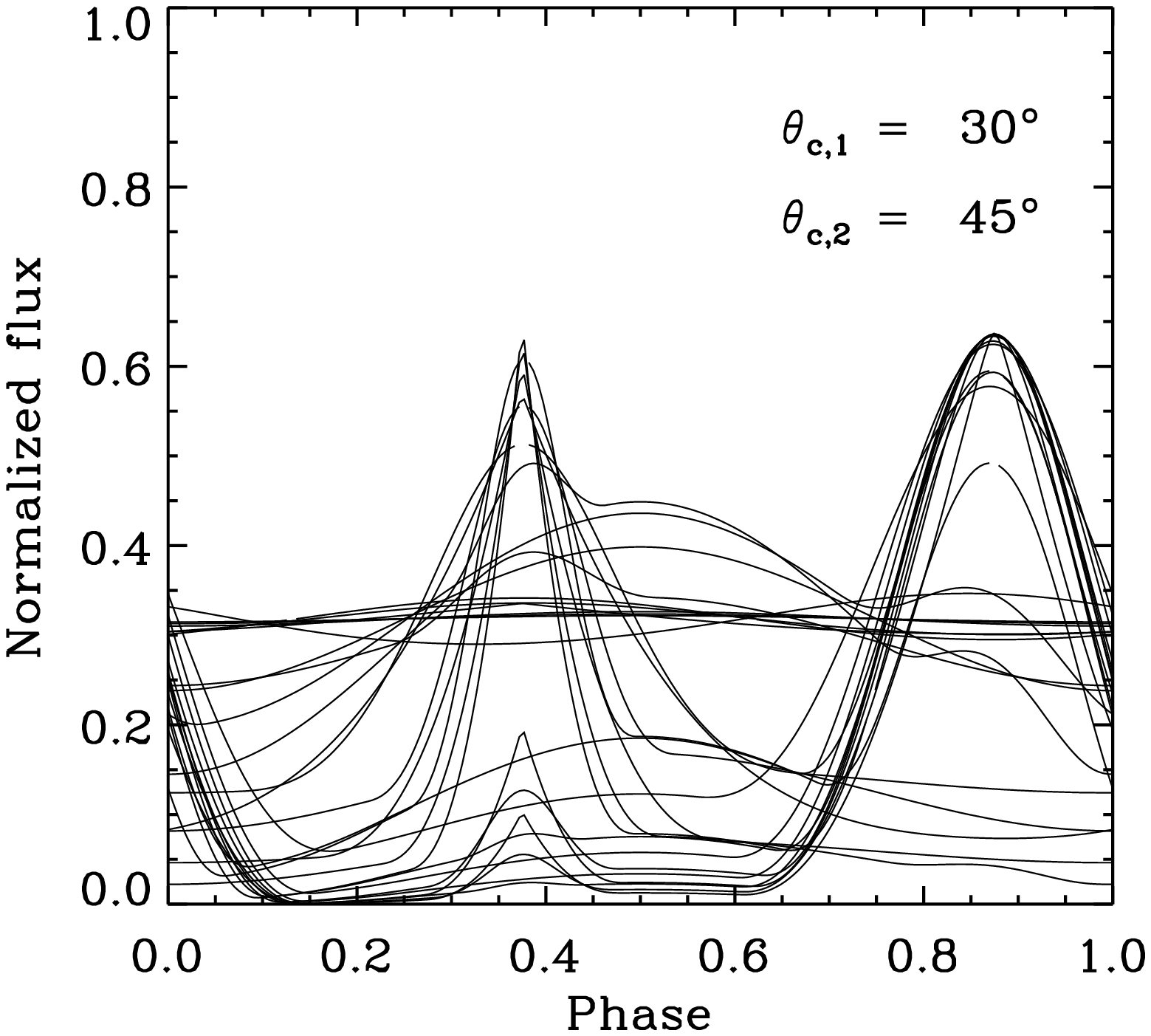} \caption{
Same as in fig. \ref{pp-one}, but for two different,
non-antipodal caps on the same meridian (\emph{left}) and with the second spot shifted in longitude by
$45^\circ$ (\emph{right}).\label{pp-mis}}
\end{figure}

\section{Discussion and conclusions}\label{discus}

In this investigation we revisited the problem of computing the pulse profiles from thermally emitting spots on the
surface of a neutron star in general relativity. Our goal has been to develop a simple approach which can be readily used
for a quantitative comparison of models with observations. \cite{belob02}, by means of a suitable approximation,
was able to derive analytical expressions for the pulse profiles in full GR for point-like, equal, antipodal spots.
However, if more realistic thermal configurations are to be accounted for, going beyond the point-like
approximation becomes necessary. We have shown that it is possible to extend Beloborodov's approach to include
(multiple) spots of finite size in different positions on the star surface. Results for the pulse profiles are
expressed by comparatively simple analytical formulas which involve only elementary functions.

A qualitative comparison between point-like and
finite-size spots is provided by Figures \ref{pp-one} (single
spot) and \ref{pp-two} (two equal, antipodal spots); since
$\theta_c=0$ produces a vanishing flux, $\theta_c=3^\circ$ was used instead
to simulate a (nearly) point-like spot (see below). Indeed, the pulse profiles in
Figure \ref{pp-two} (right) appear very similar to those
discussed by \citet[][see his fig. 4]{belob02}\footnote{The
different levels of the flat portion of the pulse profiles, or ``plateau'', are
due to our different normalization of the flux.} and the four
 ``types'' he introduced (class I, II,  III, IV) are clearly recognizable.
This is better seen in Figure \ref{pf-comp} (right), where the pulsed fraction, defined as
${\mathrm{PF}}=(F_{max}-F_{min})/(F_{max}+F_{min})$, is shown as a function of $\chi$ and $\xi$, together with
Beloborodov's analytical result (his eq. [8]). The two sets of contours are nearly indistinguishable and the maximum
pulsed fraction, equal to $(R-2R_S)/(R+2R_S)$ for point-like spots, is the same.

As expected, for larger caps the pulse shape changes, the ``plateau'' disappears and the pulsed fraction
decreases (figure
\ref{pp-two}, left). Now the constant PF contours are quite different with respect to those of a point-like spot, as
clearly shown in the left panel of figure \ref{pf-comp}. The maximal pulsed fraction is $\sim 30\%$ lower than
$(R-2R_S)/(R+2R_S)$ for $\theta_c=40^\circ$. In general, we find that the point-like approximation is reliable up
to $\theta_c\sim 5^\circ$.

\begin{figure}
\plottwo{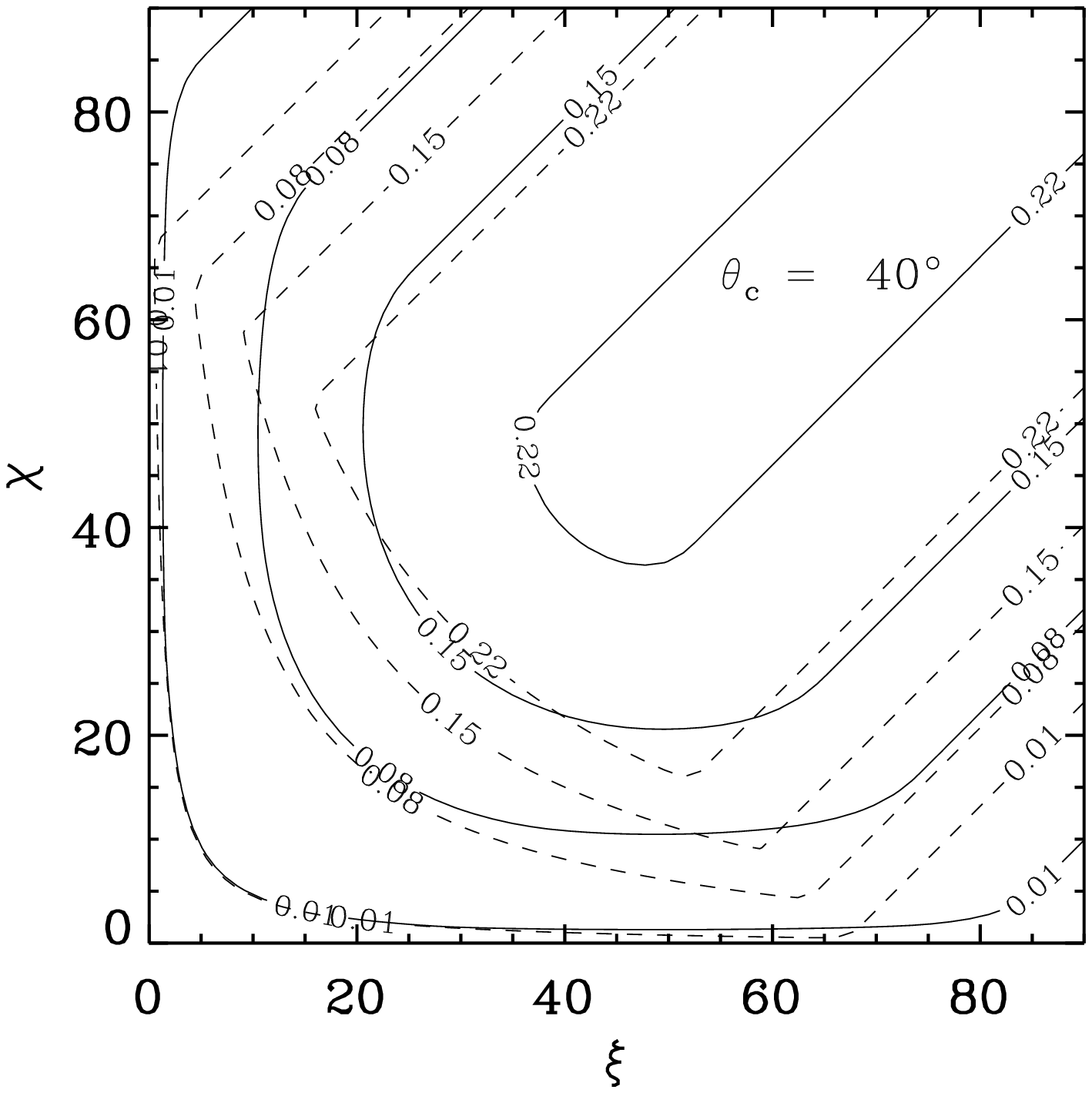}{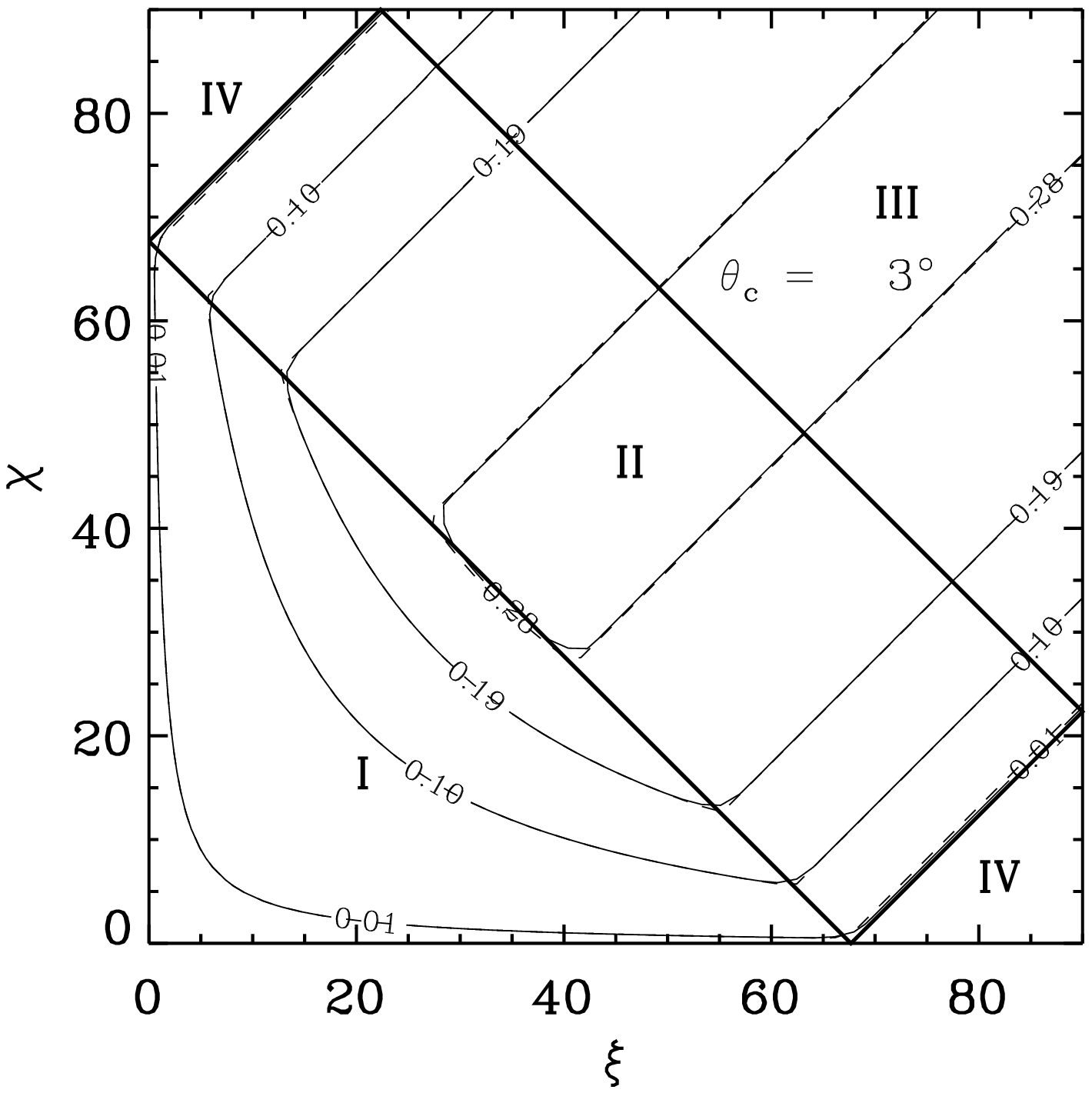} \caption{Contour plot for the pulsed
fraction as a function of the two angles $\chi$ and $\xi$ (full
lines) for two antipodal, equal spots with semi-aperture
$40^\circ$ (\emph{left}) and  $3^\circ$ (\emph{right}). The dashed
contours show the analytical result and the rectangle bounded by
the heavy lines (left panel) separates the five regions labelled
I, II, III, IV \cite[see text]{belob02}. \\\label{pf-comp}}
\end{figure}

Larger caps can be treated either using the approach described here or resorting
to methods based on general relativistic ray-tracing. We believe that the former
offers a number of advantages,
since it involves no numerical integration, and allows for a great flexibility, so that diverse
thermal configurations of the NS surface can be modeled.
An obvious limitation is that only purely blackbody
(or at any rate isotropic) emission can be treated. Despite this simple model is often successfully used in fitting
the (thermal
components) of X-ray spectra, emission from the cooling surface of isolated neutron stars is expected to
be more complicated, e.g. because the star is covered by an atmosphere, or
because the emissivity is strongly suppressed at energies below the electron plasma frequency if the
surface layers are in condensed form \cite[see e.g.][and references therein]{turolla09}. Realistic emission
models predict, to a different extent, an angular dependence of the emitted intensity. While
anisotropy is modest for non-magnetized atmospheric models \cite[][]{zavlin96}, it becomes substantial in magnetized
atmospheres \cite[][]{pavlov94}}, or in condensed surfaces \cite[][]{turolla04}. In general, it would be
impossible to compute analytically the analogues of $I_{1,2}$ (see \S\ref{single}) for a non-isotropic
radiation field. We point out, however, that in case the intensity depends on the angle $\alpha$ only, i.e.
$I_\nu=I_\nu(\theta)$ by using eq. (\ref{cosalp}), all the cosiderations presented in \S\ref{single} still hold,
although now numerical integration is required to obtain the pulse profiles. This, being the integral just over a
single variable, $\theta$,  adds only a modest complication and the present method has still advantages with
respect to fully numerical ray-tracing. Clearly this is not the case if $I_\nu$ depends on both angles, $\alpha$
and the associated azimuth, since double integrals should be evaluated.

\acknowledgements RT acknowledges financial support from an
INAF 2011 PRIN grant. We are grateful to Paolo Esposito and Silvia Zane for their helpful
comments on the manuscript.

\appendix
\section{Integrals evaluation}

An integration by parts brings the first integral into the form

\begin{equation}\label{app1}
I_1=\sin^2\theta\arccos\left[\frac{\cos\theta_c-\cos\theta_o\cos\theta}{\sin\theta_o\sin\theta}\right]
-\frac{1}{\sin\theta_o\sin\theta_c}\int\frac{\cos\theta_o-\cos\theta_c\mu}
{\sqrt{1-\displaystyle{\left(\frac{\mu-\cos\theta_o\cos\theta_c}{\sin\theta_o\sin\theta_c}\right)^2}}}\,
d\mu
\end{equation}
where $\mu=\cos\theta$. By introducing
$z=(\mu-\cos\theta_o\cos\theta_c)/(\sin\theta_o\sin\theta_c)$, the
previous expression becomes

\begin{equation}\label{app2}
I_1=\sin^2\theta\arccos\left[\frac{\cos\theta_c-\cos\theta_o\cos\theta}{\sin\theta_o\sin\theta}\right]
-\sin^2\theta_c\cos\theta_o\int\frac{dz}{\sqrt{1-z^2}}
+\cos\theta_c\sin\theta_c\sin\theta_o\int\frac{z\,
dz}{\sqrt{1-z^2}}\,,
\end{equation}
which, after some trivial manipulations, yields eq. (\ref{integpindex}).

$I_2$ is handled in a similar way. After integrating by parts, one
gets

\begin{equation}\label{app3}
I_2=-2\cos\theta\arccos\left[\frac{\cos\theta_c-\cos\theta_o\cos\theta}{\sin\theta_o\sin\theta}\right]
+\frac{2}{\sin\theta_o\sin\theta_c}\int\frac{\mu(\cos\theta_o-\cos\theta_c\mu)}{1-\mu^2}\frac{d\mu}
{\sqrt{1-\displaystyle{\left(\frac{\mu-\cos\theta_o\cos\theta_c}{\sin\theta_o\sin\theta_c}\right)^2}}}\,.
\end{equation}
Upon writing

\begin{equation}\label{app4}
\frac{\mu(\cos\theta_o-\cos\theta_c\mu)}{1-\mu^2}=\cos\theta_c+\frac{1}{2}\left(-\frac{\cos\theta_o+\cos\theta_c}{1+\mu}+
\frac{\cos\theta_o-\cos\theta_c}{1-\mu}\right)\,,
\end{equation}
eq. (\ref{app3}) can be cast as

\begin{eqnarray}\label{app5}
I_2&=&-2\cos\theta\arccos\left[\frac{\cos\theta_c-\cos\theta_o\cos\theta}{\sin\theta_o\sin\theta}\right]
+2\cos\theta_c\int\frac{dz}{\sqrt{1-z^2}}\cr
&&-\left(\cos\theta_o+\cos\theta_c\right)\int\frac{d\mu}{(\mu+1)\sqrt{-\mu^2+2\cos\theta_c\cos\theta_o\mu-\cos^2\theta_c
-\cos^2\theta_o+1}}\cr
&&-\left(\cos\theta_o-\cos\theta_c\right)\int\frac{d\mu}{(\mu-1)\sqrt{-\mu^2+2\cos\theta_c\cos\theta_o\mu-\cos^2\theta_c
-\cos^2\theta_o+1}}\,.
\end{eqnarray}
The last two integrals in eq. (\ref{app5}) are of the general type

\begin{equation}\label{app6}
\int\frac{dx}{(x+p)\sqrt{ax^2+bx+c}}=\frac{1}{\sqrt{bp-ap^2-c}}\arcsin\frac{(b-2ap)x-bp+2c}{(x+p)\sqrt{b^2-4ac}}
\end{equation}
($ap^2-bp+c <0\,,b^2-4ac > 0\,;$ e.g. \citealt{pbm92}). Note,
however, that the previous expression is valid only if $x+p>0$. If
$x+p<0$, as in the last integral in eq. (\ref{app5}) where $\mu-1
<0$, a minus sign must be placed in front of the result. Making
use of eq. (\ref{app6}) and after some algebra, eq.
(\ref{integsindex}) is finally recovered.

\end{document}